\documentclass[aps,prl,twocolumn,superscriptaddress,showpacs,10pt]{revtex4-1}

\usepackage{graphicx}
\usepackage{mathtools}
\usepackage{amssymb,amsmath}
\usepackage[caption = false]{subfig}
\usepackage{dsfont}
\usepackage{booktabs}
\usepackage{placeins}
\usepackage{verbatim}
\usepackage{comment}
\usepackage{color}
\usepackage{float}
\usepackage{multirow}
\usepackage[utf8]{inputenc}
\usepackage[export]{adjustbox}
\usepackage{hyperref}
\hypersetup{colorlinks=false}
\usepackage{lineno}
\usepackage{upgreek}

\newcommand{\reffig}[1]{Fig.~\ref{fig:#1}}

\newcommand{\refeqs}[1]{Eq.~\eqref{eq:#1}}

\newcommand{\lkedit}[1]{\textcolor{black}{#1}}
\newcommand{\td}{\,\mathrm{d}}
\newcommand{\bfV}{\mathbf{V}}
\newcommand{\bfU}{\mathbf{U}}
\newcommand{\bff}{\mathbf{f}}

\newcommand{\etal}{\textit{et al.~}}

\DeclareMathOperator*{\argmin}{argmin}

\begin{document}
 

\title{Data-driven detection of drifting system parameters}

\date{\today}

\author{Logan M. Kageorge}
\affiliation{School of Physics, Georgia Institute of Technology, Atlanta, Georgia 30332, USA}
\author{Roman O. Grigoriev}
\affiliation{School of Physics, Georgia Institute of Technology, Atlanta, Georgia 30332, USA}
\author{Michael F. Schatz*}
\affiliation{School of Physics, Georgia Institute of Technology, Atlanta, Georgia 30332, USA}
\affiliation{*email: michael.schatz@physics.gatech.edu}

\maketitle


\section{Abstract}
Data taken from observations of the natural world or laboratory measurements often depend on parameters which can vary in unexpected ways.
In this paper we demonstrate how machine learning can be leveraged to detect changes in global parameters from variations in an identified model \lkedit{using only observational data. 
This capability, when paired with first principles analysis, can effectively distinguish the effects of these changing parameters from the intrinsic complexity of the system.}
Here we illustrate this by identifying a set of governing equations for an experiment generating a weakly turbulent fluid flow, then analyzing variation in the coefficients of these equations to unravel the drift in its physical parameters. 

\section{Introduction}
In the last few decades, machine learning has emerged as a viable means for overcoming obstacles spanning many areas thanks to recent advances in data acquisition, storage, and computing power. Machine learning can even outperform humans, providing new capabilities for solving problems with cumbersome amounts of data, such as voice activity detection \cite{shin_2010}, search engine optimization \cite{boyan_1996}, financial fraud detection \cite{perols_2010}, and protein folding and dynamics \cite{noe_2020}. 

In the context of the physical sciences, the methods and results of machine learning take many different forms. Reinforcement learning can be used to optimize experimental conditions and improve chemical reaction outcomes \cite{zhou_2017}, deep learning algorithms can be trained to produce fluid flow visualization or speed up computations \cite{raissi_2020, kochkov_2021}, and sparse regression can discover nonlinear models governing biological network formation and growth \cite{mangan_2016}. While many of these studies have focused on purely data-driven methods to formulate equation-based models \cite{schmidt_2009, bongard2007}, recent hybrid approaches combine machine learning techniques with physical principles to more robustly and efficiently identify algebraic or partial differential equation models from data \cite{reinbold_2021, gurevich_2019, gurevich_2021}. 

Real-world systems both deterministic and stochastic are frequently subject to drifting global parameters, which can have drastic consequences \cite{zeeman_1976, thom_1972, kaszas_2019, scheffer_2009}. Parameter drift plays an important role in settings at every scale, from rising greenhouse gas emissions in climate dynamics \cite{lenton_2011,ghil_2013} to thermal fluctuations in quantum systems \cite{cortez_2017}. These drifts are traditionally detected indirectly from observations of qualitative changes in a system's dynamics, however these changes are often the result of multiple drifting parameters which can be difficult to distinguish. Moreover, different sets of parameters can yield qualitatively similar dynamics, obfuscating any drift that may be occurring. Thus, the need for a more direct, quantitative method of identifying parameter drift is necessary to accurately model systems with complex dynamics.

We present here a general method for detecting parameter drift, and demonstrate its efficacy in the context of a fluid flow in a laboratory setting.
Such experiments are subject to a wide variety of complications that can lead to drifting parameters and thus spurious data. 
We utilize sparse physics-informed discovery of empirical relations (SPIDER), a machine learning approach that has proven to be particularly adept at synthesizing mathematical models from noisy, incomplete, high-dimensional simulated and experimental data \cite{gurevich_2021,reinbold_2020,reinbold_2021}, to identify a governing equation for a weakly turbulent fluid experiment. 
We then vary the global parameters to simulate common experimental complications and to observe the effect on the identified model terms and their coefficients. 
Finally, we compare the model and the functional dependence of these coefficients to the predictions from first principles analysis \cite{suri_2014}, and show that the coefficients alone can serve as a benchmark for troubleshooting and detecting errors in the experimental setup.

\begin{figure*}[htb]
    \centering
    \subfloat[]{\includegraphics[width=0.3\textwidth, valign=c]{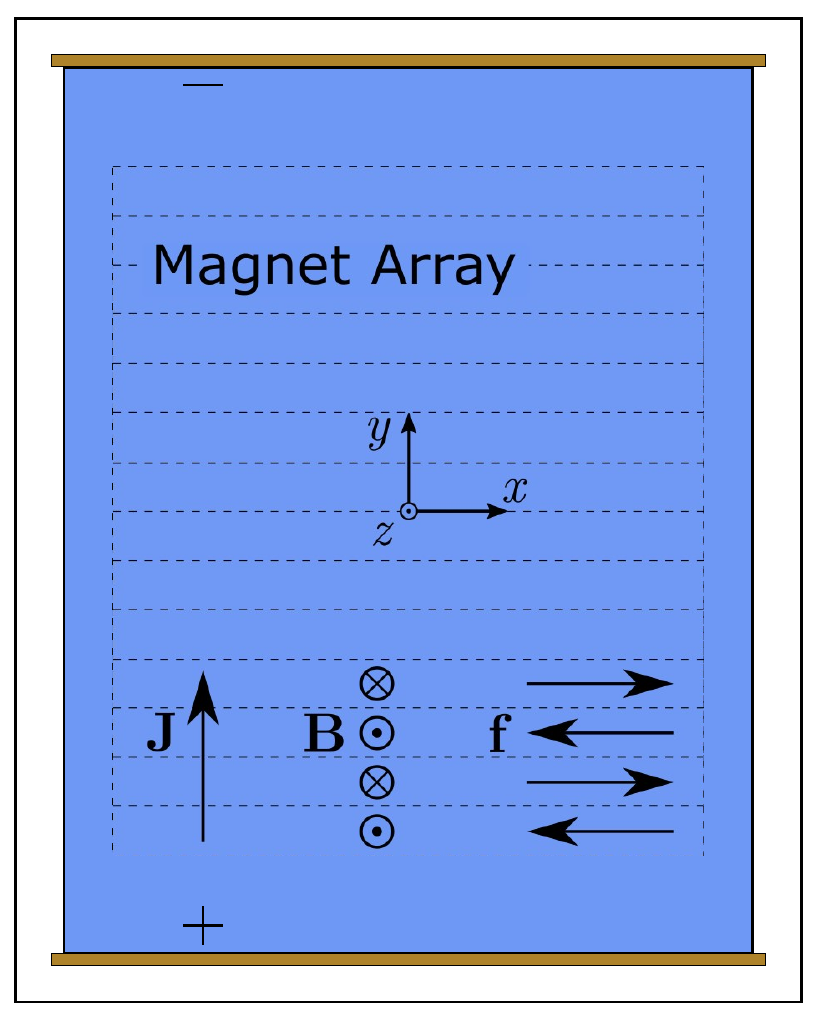}} \qquad
    \subfloat[]{\includegraphics[width=0.3\textwidth, valign=c]{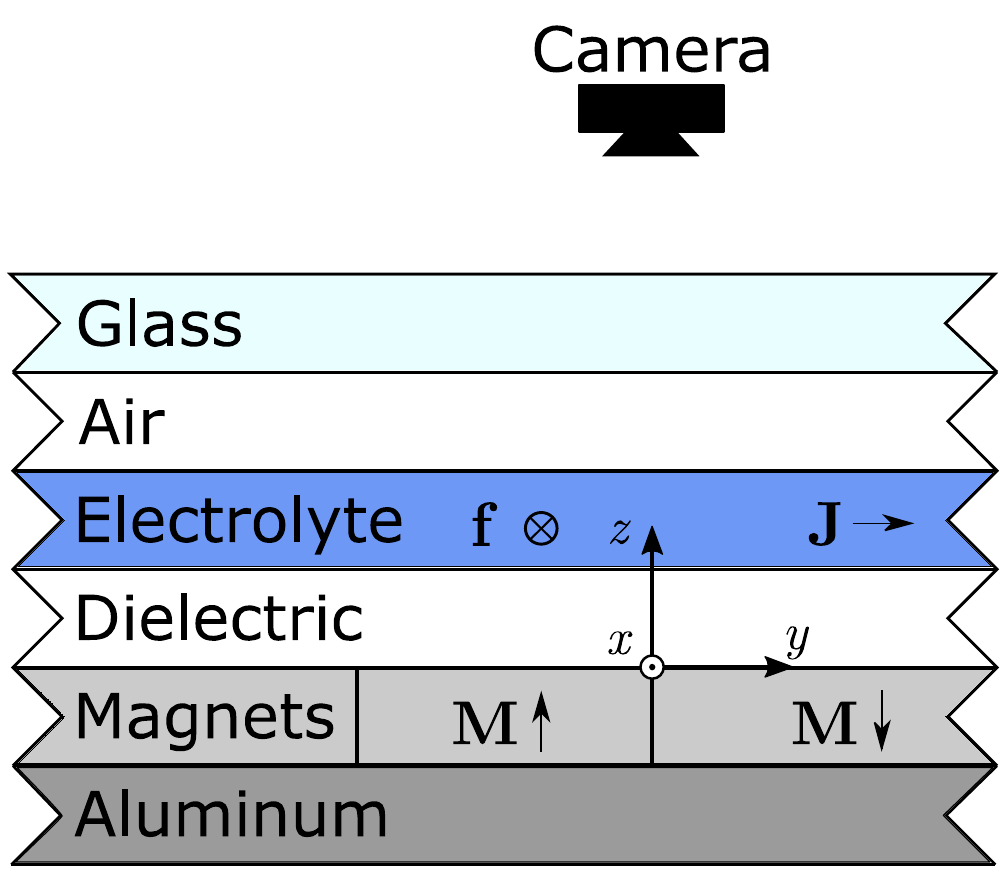}} \qquad
    \subfloat[]{\includegraphics[width=0.3\textwidth, valign=c]{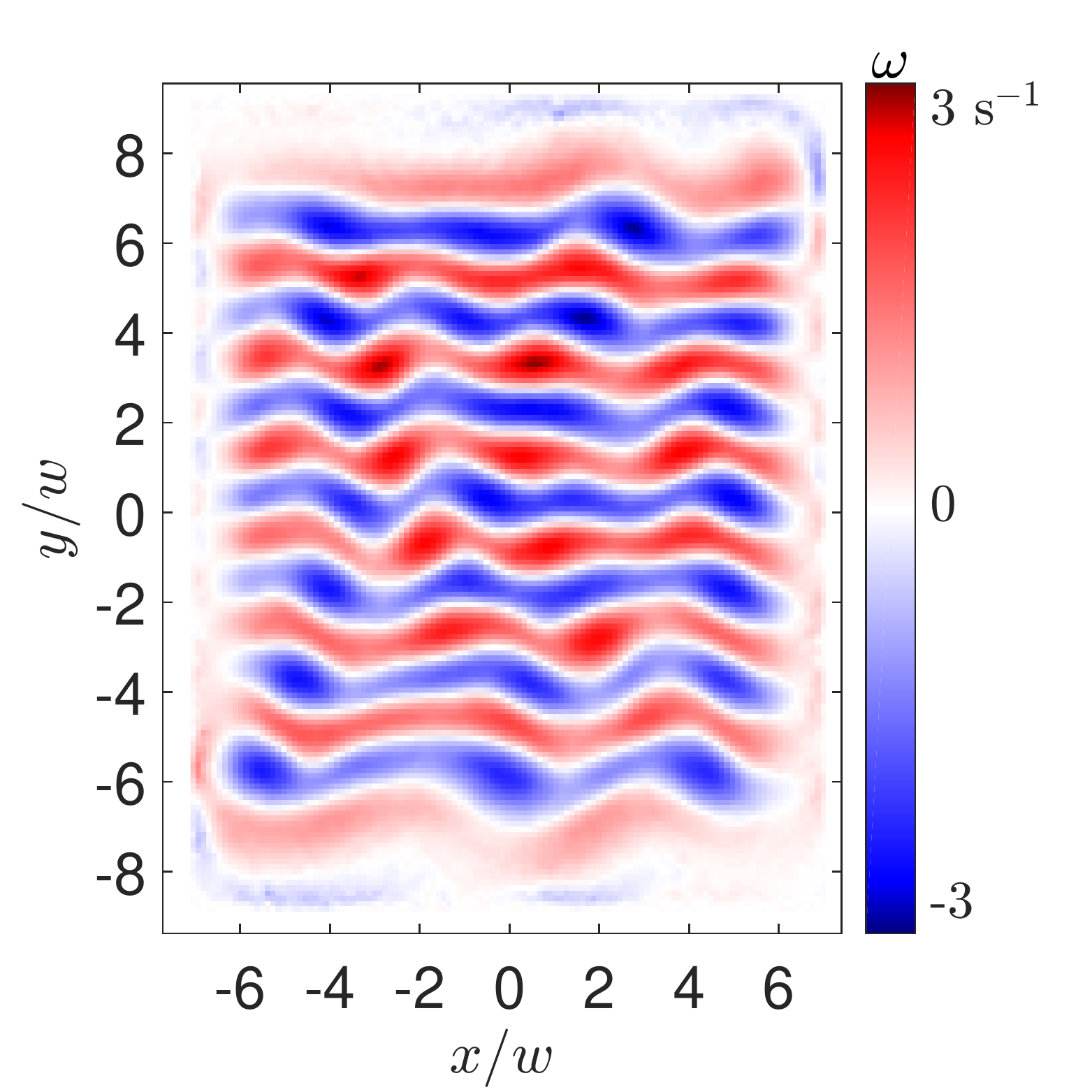}}
    \caption{{\bf Experimental Schematic.} The experimental setup top (a) and side (b) view demonstrate the geometry and forcing for the thin electrolyte fluid flow studied here. Permanent magnets below the fluid generate a magnetic field $\mathbf{B}$ while a current density $\mathbf{J}$ is driven between the two electrodes, creating a Lorentz forcing profile $\bff$ that is nearly sinusoidal. The dielectric layer allows the electrolyte to overcome the no-slip boundary at the bottom of the apparatus, creating a quasi-two-dimensional flow. The working temperature is maintained by a water reservoir below the aluminum base plate, while the glass lid prevents evaporation. A heatmap of the fluid vorticity field (c) shows a typical snapshot of the weakly turbulent flow at Reynold's number Re $\approx 21$.}
    \label{fig:setup}
\end{figure*}

The fluid flow is an approximation of two-dimensional (2D) Kolmogorov flow, a shear flow driven by time-independent forcing that is sinusoidal in space which develops patterns of vortices that exhibit spatiotemporal chaos.
It is generated in a shallow immiscible fluid bi-layer of dielectric and electrolyte, and is driven by the Lorentz interaction of a direct current density and the field of a permanent magnet array (See \reffig{setup} and the Methods section for more details). 

\lkedit{The dynamics featured by this system are highly dependent on a number of global parameters. 
The fluids themselves have material parameters which are dependent on a number of environmental factors, as well as geometric parameters such as the shape of the container. 
This study will focus primarily on parameters that are likely to drift while performing an experiment, namely the viscosity, density, and depths of the two fluids.}

While running the fluid experiment, there are many physical processes \lkedit{which are difficult to detect in the laboratory and can lead to a drift in these parameters}. The fluids are contained in a rectangular geometry that must be assembled from pieces and thus may not be perfectly sealed, \lkedit{causing fluids to slowly leak which decreases their depth. Moreover, the dielectric used is volatile and has a low surface tension and viscosity, so any leaked dielectric evaporates before it can be detected.} 
The electrolyte under study is highly viscous and experiences Joule heating as a current is run through it. 
\lkedit{The resulting temperature change can cause both fluids' densities and viscosities to vary, and leaves the electrolyte prone to evaporation, which decreases its depth.}

In this study, we incrementally change the depth of each fluid layer and the temperature to mimic these typical experimental complications. 
Throughout these trials, measures are taken to ensure that all other parameters such as Reynolds number are kept constant so as to only measure the effects of varying one parameter at a time.
We expect that SPIDER will find a parsimonious model that can capture the fluid's dynamics in different parameter regimes using only a reconstructed time series of the horizontal components of the fluid's velocity field, as has been previously reported \cite{reinbold_2021}.

\section{Results}
In every trial performed, SPIDER returned the parsimonious model \ref{eq:Q2DNS}. This same model has been recovered previously at various Reynolds number (Re) \cite{reinbold_2021}, and is unchanged at every depth of each fluid and at every temperature tested with a relatively low residual, $\eta \lesssim 0.02$ (see the Supplementary Notes for residual details). As predicted, this model is very similar in appearance to the Navier-Stokes equation, but with a few modifications to account for the vertical confinement that allows for 2D modeling of an inherently 3D system. Namely, we see a coefficient $c_A < 1$ on the advection term $c_A\bfU\cdot \nabla \bfU$ and a Rayleigh friction term $c_F \bfU$. 
The former accounts for reduced advection due to the interaction with the bottom boundary, while the latter describes linear drag from the bottom, and was first proposed by Bondarenko \etal to describe a similar flow \cite{bondarenko_1979}.
The viscous term $c_V \nabla^2 \bfU$ has a coefficient $c_V$ that plays the role of viscosity, but lies somewhere between the viscosities of the two liquids. 
Without a direct measurement of the pressure or forcing fields, the coefficients $c_f$ and $c_p$ are only defined up to a multiplicative scale, and are therefore not useful as quantitative diagnostic tools.
\lkedit{Note that in many disciplines, $c_A$, $c_V$, and $c_F$ may be called model parameters, however their dependences on the geometric and material parameters of the system do not have a straightforward interpretation.
To avoid terminology that would confuse the system parameters with the observed quantities, these three numbers will be referred to only as coefficients for the remainder of this manuscript.}
Throughout the trials, it is only $c_A$, $c_V$, and $c_F$ that vary with the global parameters, so we describe here the qualitative trends in these coefficients. 

\begin{figure*}[htb]
    \centering
    \subfloat[]{\includegraphics[width=0.33\textwidth]{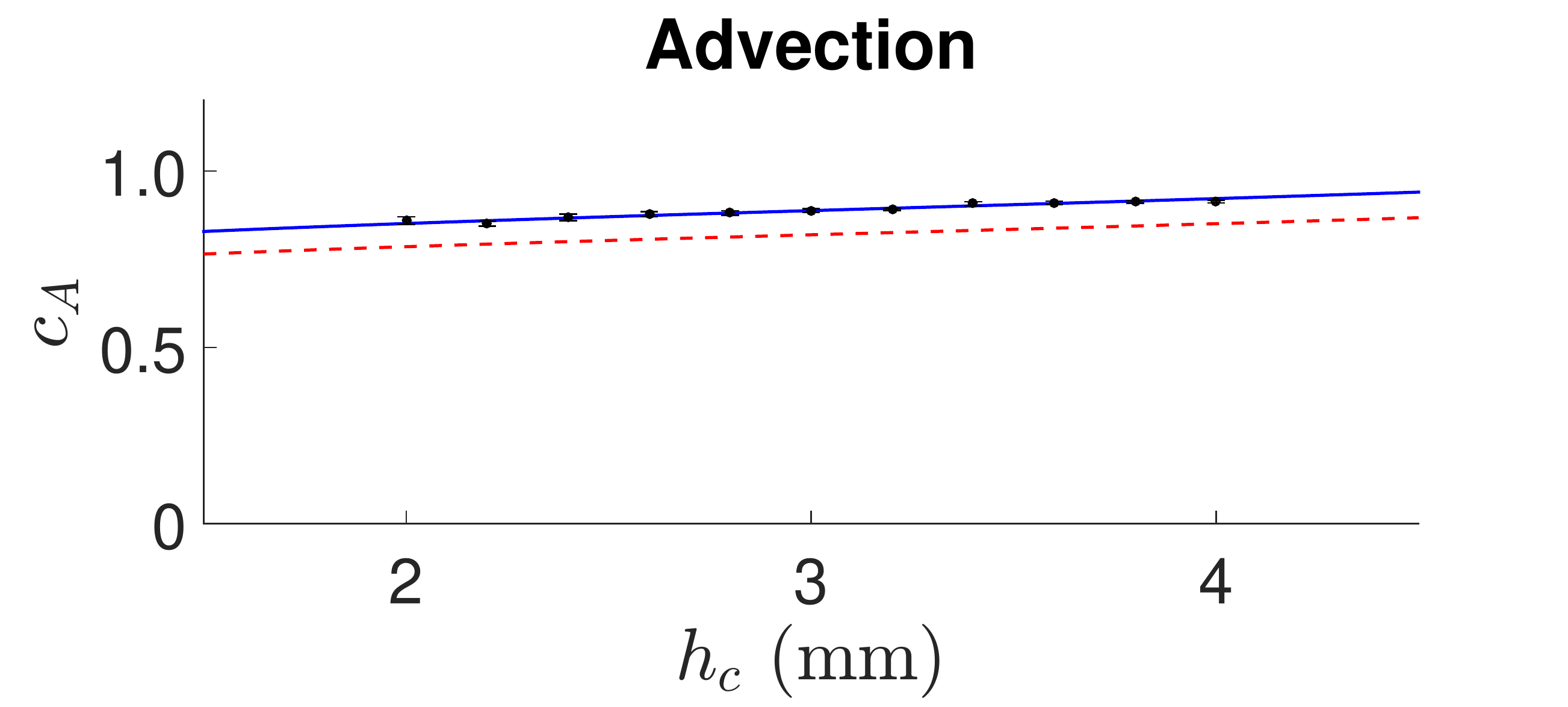}}
    \subfloat[]{\includegraphics[width=0.33\textwidth]{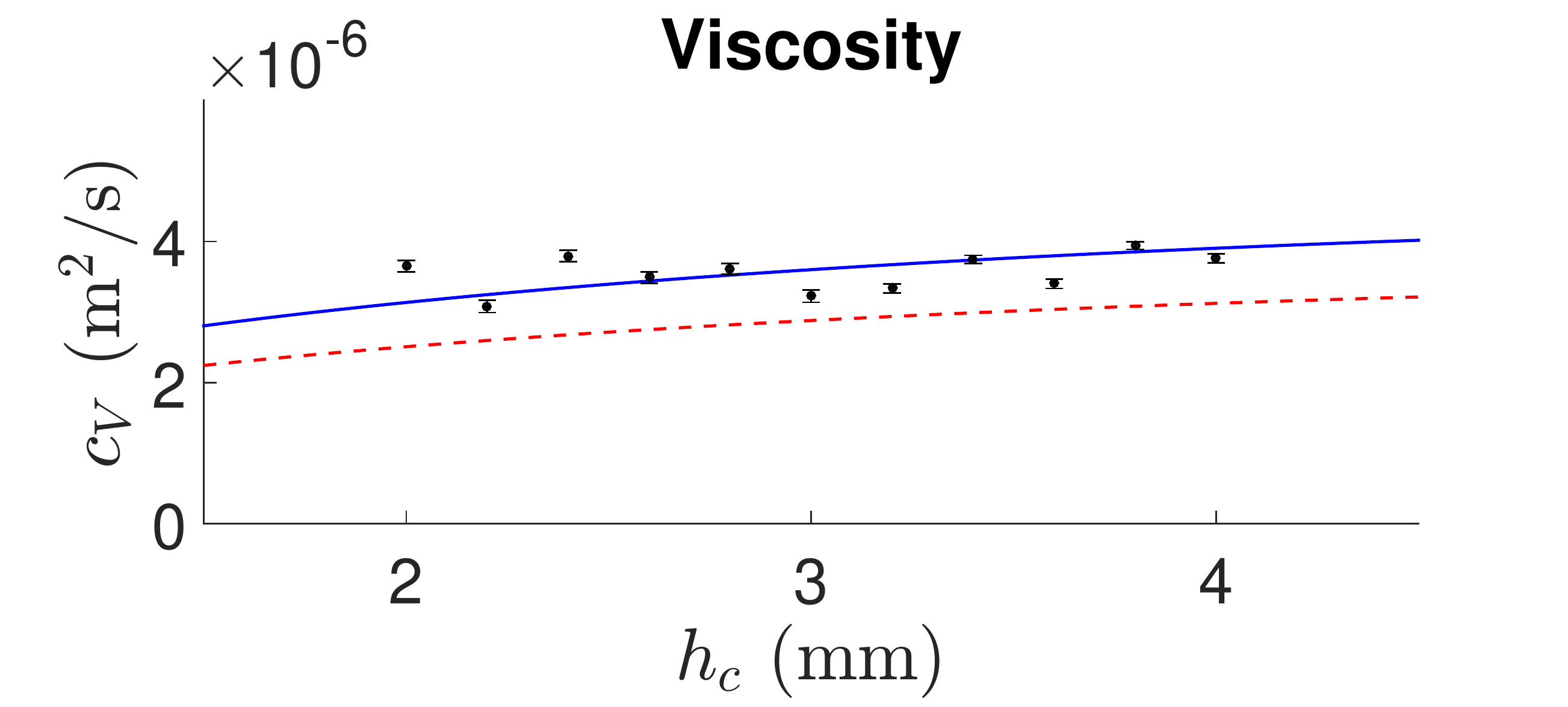}}
    \subfloat[]{\includegraphics[width=0.33\textwidth]{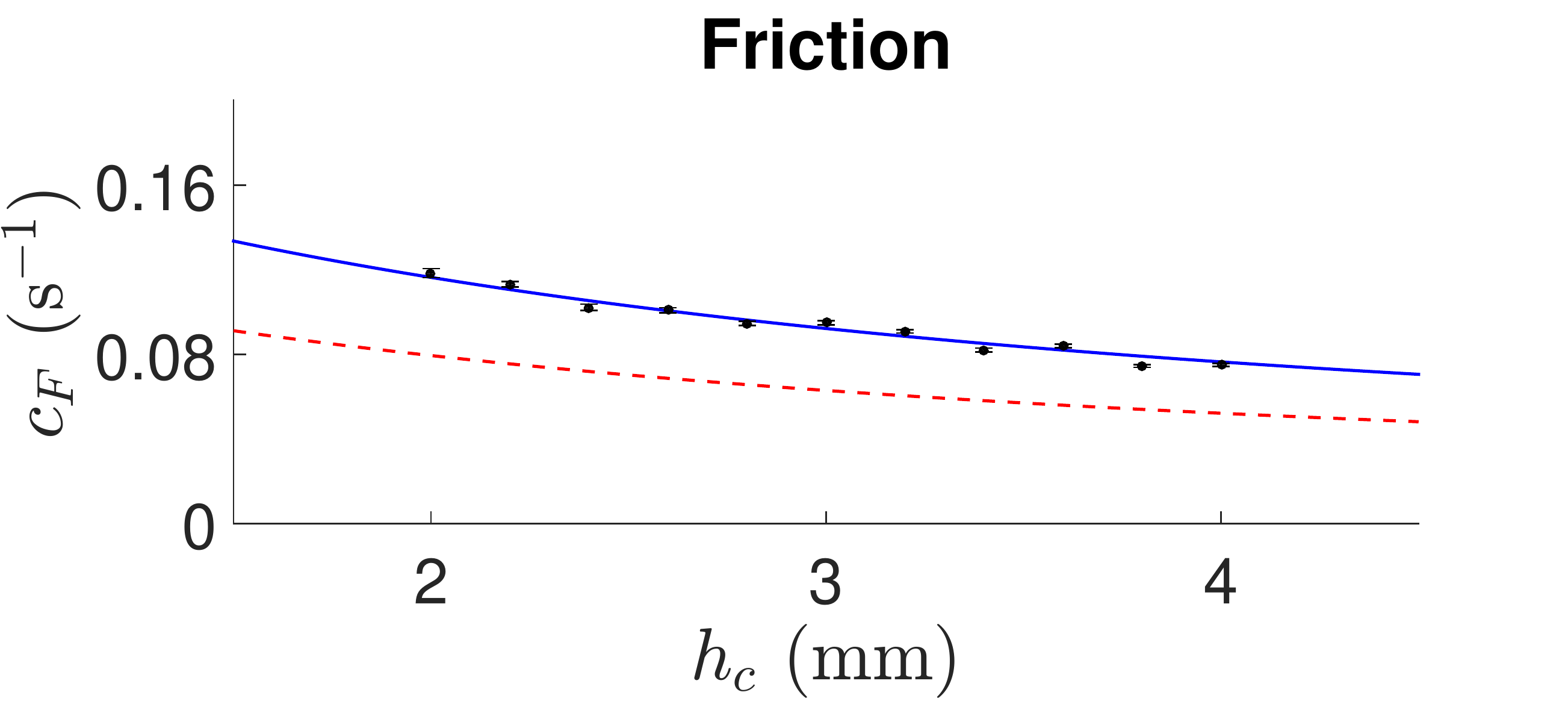}}
    \caption{{\bf SPIDER results: varying electrolyte depth.} Model coefficients (a) $c_A$, (b) $c_V$, and (c) $c_F$ of the modified 2D Navier-Stokes, \refeqs{Q2DNS} as determined by SPIDER as the electrolyte depth $h_c$ is varied. Increasing $h_c$ tends to decrease the linear friction coefficient $c_F$, while viscosity $c_V$ is increased, Advection $c_A$ also rises, though only slightly. These trends agree with predictions from first principles analysis, shown as a dotted red line. Rescaling these predictions via least squares fitting gives a nice trend line for the SPIDER data, as shown in blue. Error bars represent standard deviations.}
    \label{fig:c_hc}
\end{figure*}

In the first trial, we systematically vary the depth of the electrolyte layer $h_c$ to simulate difficulties in ensuring the depth of the fluid layer.
As the electrolyte depth increases, we observe a minuscule increase in the advection coefficient $c_A$. We also see an increase in the viscous term $c_V$; it is likely that this coefficient trends toward the viscosity value of whichever fluid is present in larger quantity, in this case the highly-viscous electrolyte. We observe a much more noticeable decrease in the friction term $c_F$, as the bottom boundary plays a less substantial role in the dynamics of the fluid. These results are plotted as black symbols in \reffig{c_hc}.

\begin{figure*}[htb]
    \centering
    \subfloat[]{\includegraphics[width=0.33\textwidth]{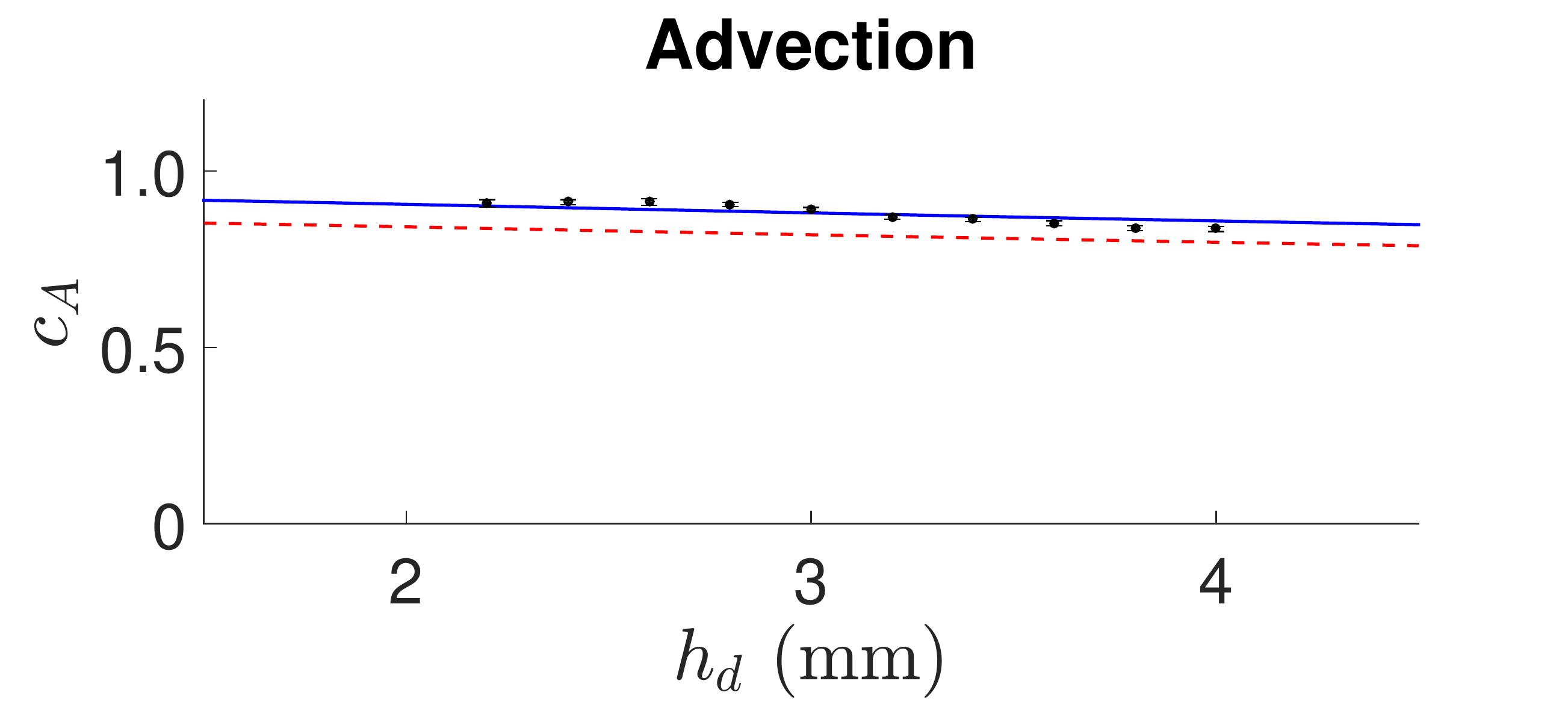}}
    \subfloat[]{\includegraphics[width=0.33\textwidth]{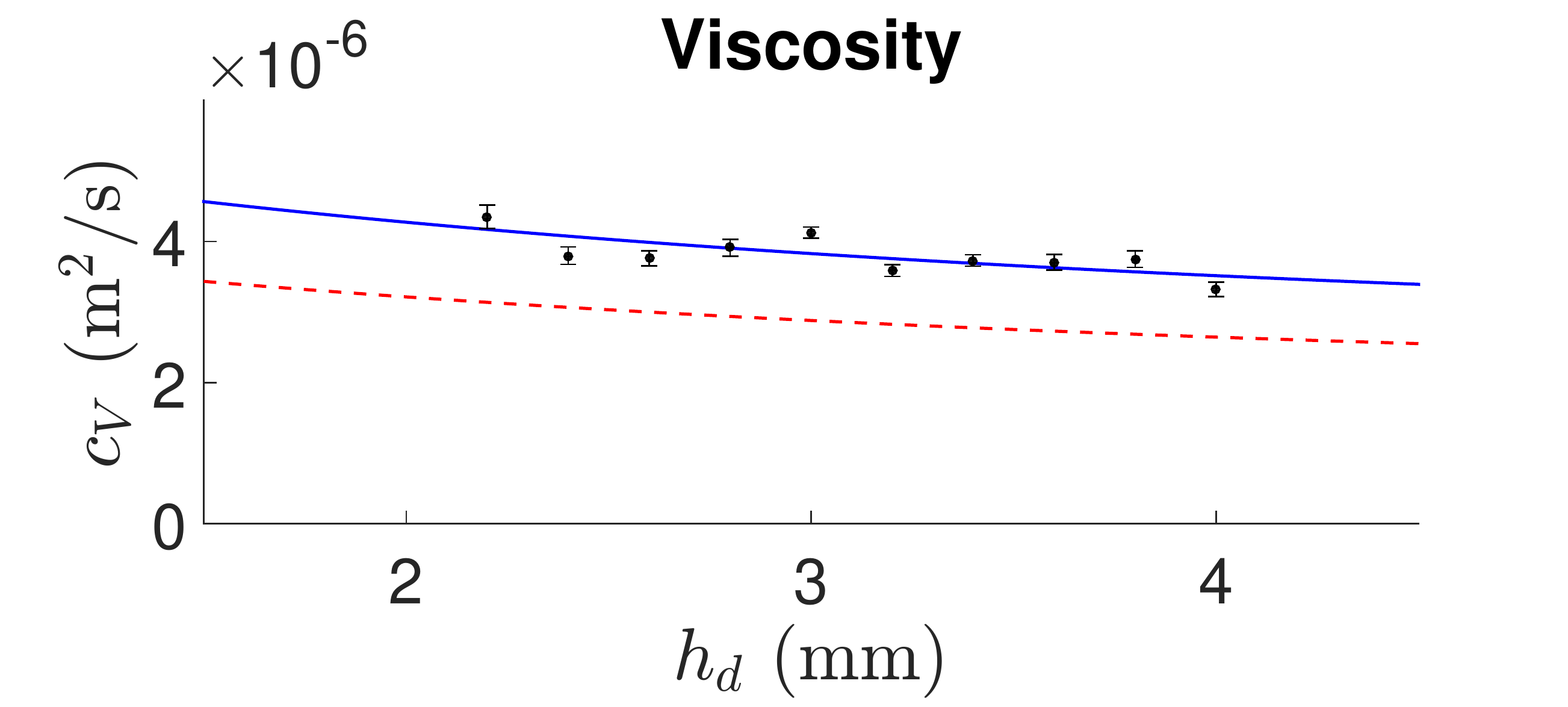}}
    \subfloat[]{\includegraphics[width=0.33\textwidth]{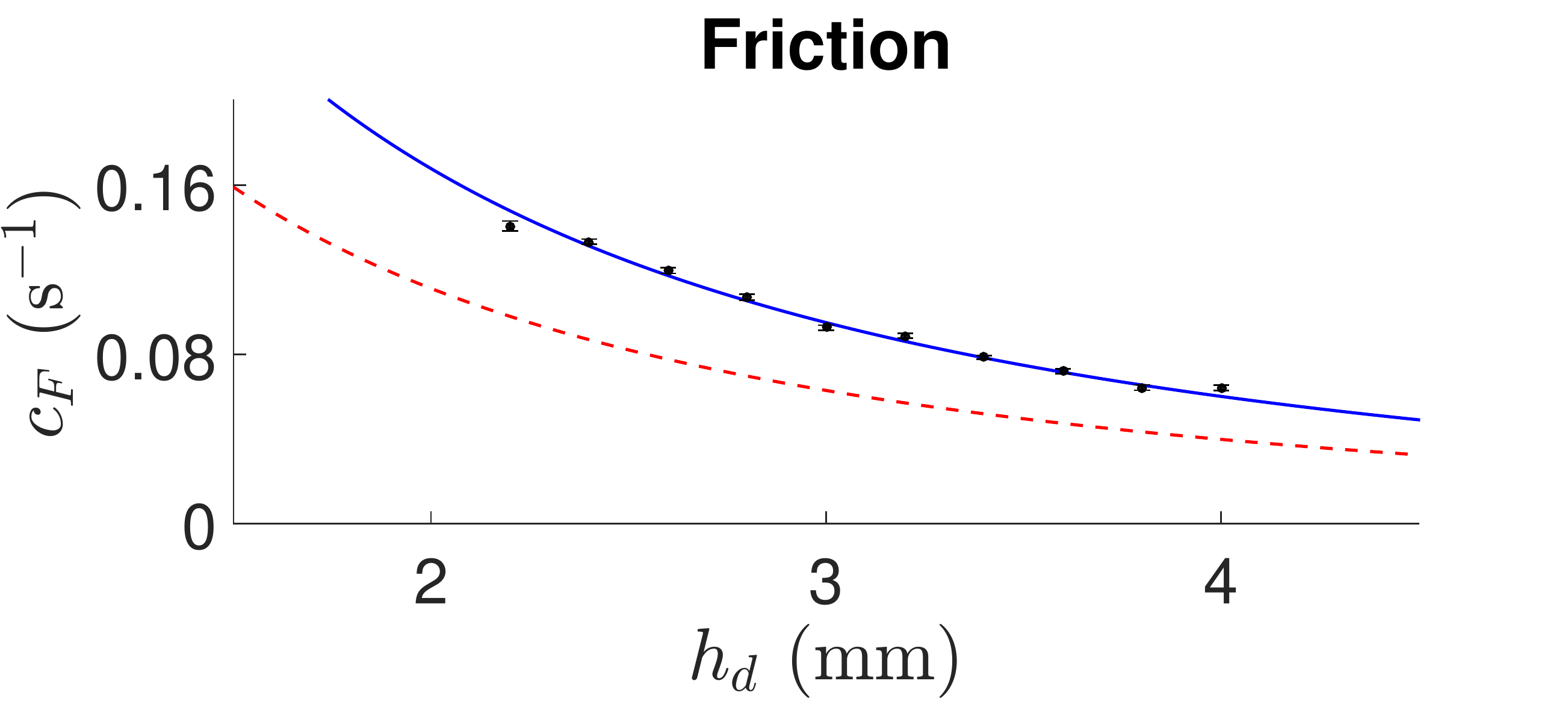}}
    \caption{{\bf SPIDER results: varying dielectric depth.} As the dielectric depth $h_d$ is varied, SPIDER again returns \refeqs{Q2DNS} as a parsimonious model, with coefficients (a) $c_A$, (b) $c_V$, and (c) $c_F$. Increasing $h_d$ marginally decreases the advection term $c_A$, the viscous term $c_V$ drops a bit more, and friction $c_F$ sharply decreases. These are once again well-predicted by the first principles model, and follow the rescaled predictions extremely well. Error bars represent standard deviations.}
    \label{fig:c_hd}
\end{figure*}

In the second trial, the dielectric depth $h_d$ is varied to replicate fluid losses from leaks. SPIDER again returns \refeqs{Q2DNS} as the governing equation, but with different trends in the coefficients. As dielectric depth increases, $c_A$ decreases only marginally. $c_V$ decreases now, trending toward the lower viscosity value of the dielectric. $c_F$ decreases sharply, as increasing the dielectric height moves the driven electrolyte layer further away from the bottom boundary, decreasing friction. The black symbols in \reffig{c_hd} plot these trends.

\begin{figure*}[htb]
    \centering
    \subfloat[]{\includegraphics[width=0.33\textwidth]{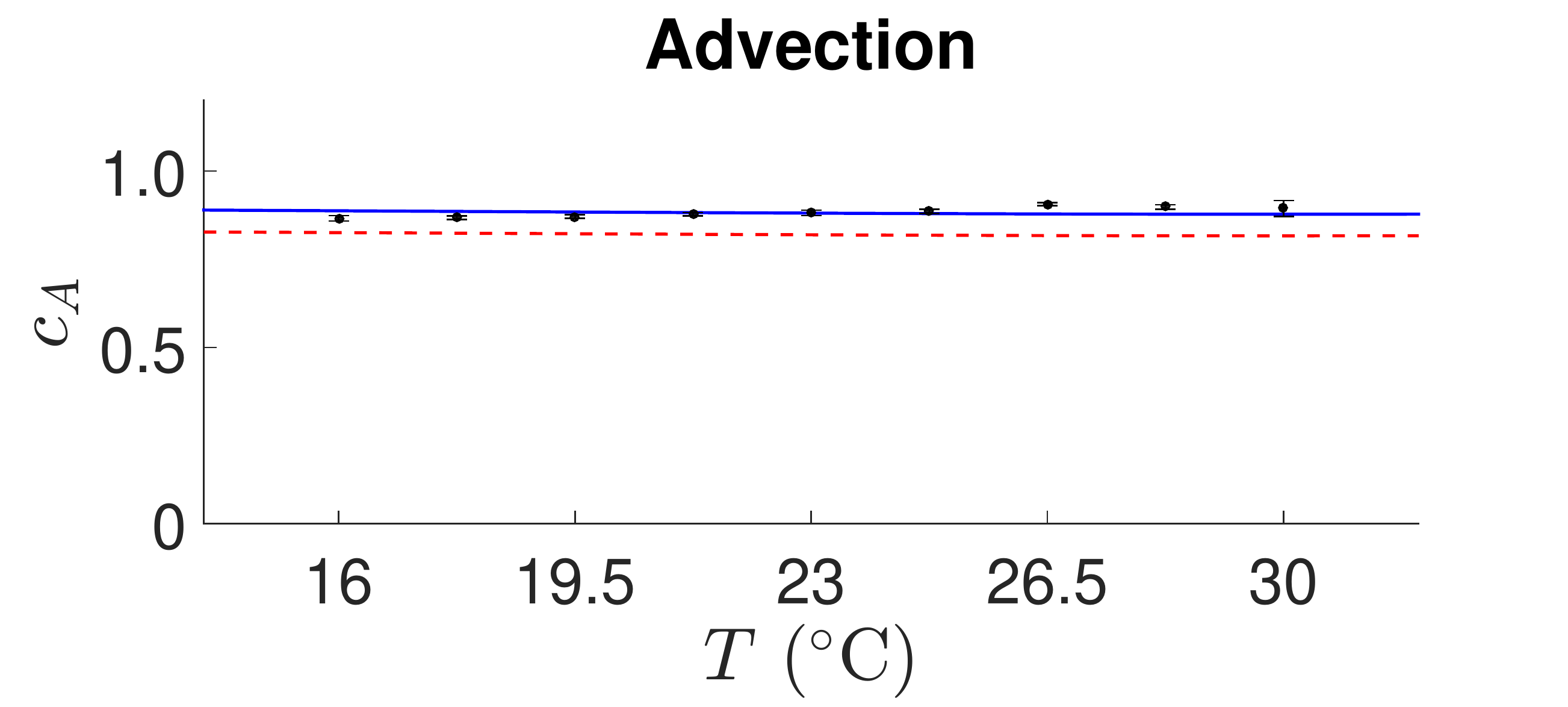}}
    \subfloat[]{\includegraphics[width=0.33\textwidth]{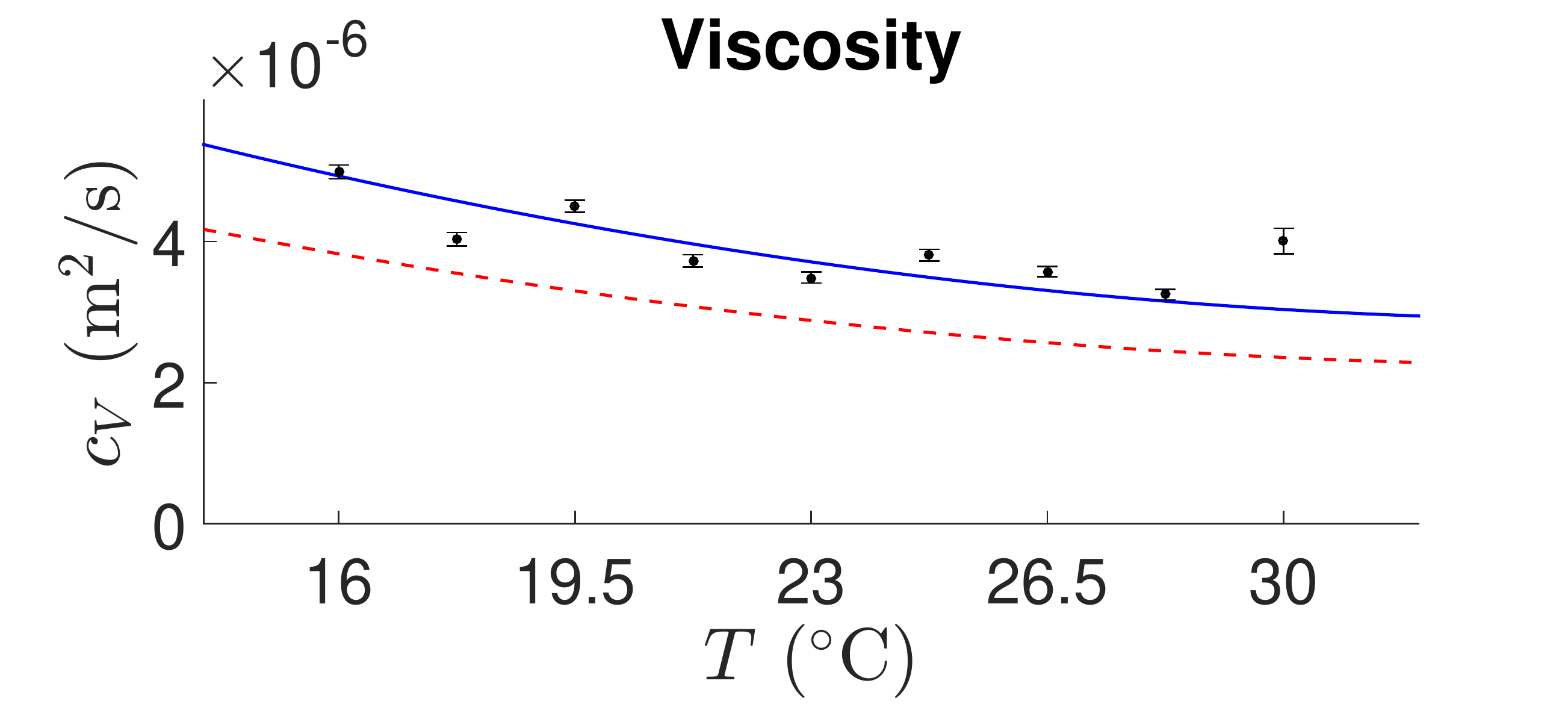}}
    \subfloat[]{\includegraphics[width=0.33\textwidth]{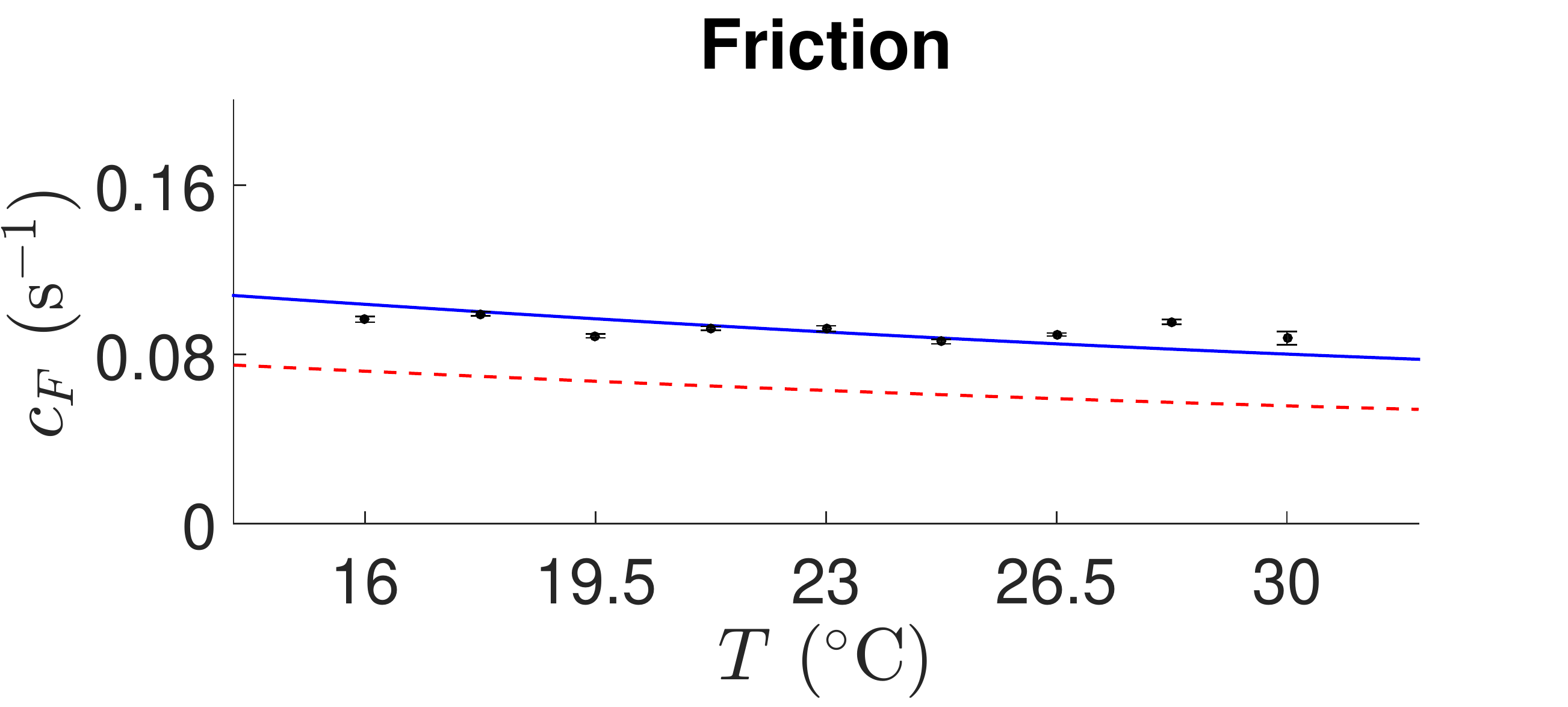}}
    \caption{{\bf SPIDER results: varying temperature.} Varying temperature $T$ gives new trends for \refeqs{Q2DNS} coefficients (a) $c_A$, (b) $c_V$, and (c) $c_F$. Increasing $T$ does not have much effect on the advection or friction terms $c_A$ and $c_F$, but the viscous coefficient $c_V$ decreases, as shown by the dotted trend lines. The first principles model predictions are again shown as a dotted red line, with a solid blue to show the rescaled predictions. The SPIDER result for $c_V$ at $T$ = 30$^\circ$C may be seen as an outlier; \lkedit{including it in the least-squares scaling procedure produces a $c_2$ fit curve slightly higher than the one shown, and does not significantly affect the scaling of the other two coefficients}. Error bars represent standard deviations.}
    \label{fig:c_T}
\end{figure*}

We vary temperature $T$ in the third trial to simulate Joule and viscous heating that can occur if thermal regulation fails. \reffig{c_T} shows the new trend in the coefficients of \refeqs{Q2DNS}; in these trials we see minimal variation in the advection coefficient $c_A$ or the friction $c_F$. However, the viscous term $c_V$ drops dramatically as temperature rises, reflecting the inverse relationship of viscosity and temperature that is typical in liquids (see the Supplementary Notes for details). 

It should be noted that at the highest temperature tested (30$^\circ$C), water evaporates out of the electrolyte fast enough to fog the sealing glass plate, preventing imaging. For this experiment, the plate was removed for the data set after the set temperature was achieved, which resulted in an outlier in the data. Most notably, the viscous coefficient at this temperature is much higher than the trend and has a larger standard deviation. Including this data point does not significantly change the results presented here, nevertheless it has been omitted from the computations in the remainder of the paper.

\section{Discussion}

As the three global parameters are changed, even though Re is kept fixed, the dynamics of the flow are quite different at the extreme ends of of each parameter regime. The evolution of the flow is much faster and more chaotic at high temperature, likely because there is less viscous diffusion. Likewise, thicker layers of electrolyte and dielectric both speed up the evolution of the flow, as vertical diffusion or friction plays a smaller role in the dynamics. \lkedit{This description alone shows that three different global parameters can have similar effects in the flow's evolution. Additionally, many systems naturally exhibit a wide variety of dynamic behaviors as they evolve. Bursting and other intermittent phenomena are common features in turbulent fluids which appear as moments of accelerated evolution \cite{rao_1971, frisch_1985, frisch_1977, itano_2001}, and could easily be mistaken for the kind of complex behavioral shift attributed to parameters drift. Clearly a qualitative description of the flow's behavior does not provide nearly enough information to diagnose drifting parameters, and therefore our machine learning approach provides a compelling alternative.}

The most notable feature of the results shown in \reffig{c_hc}-\ref{fig:c_T} is the fact that each of the system parameters tested has a unique signature in the monotonic trends of the model coefficients. In other words, we only see $c_V$ decrease while $c_F$ remains constant when temperature increases; $c_V$ and $c_F$ both increase only when the dielectric depth decreases; and $c_V$ increases while $c_F$ decreases only when the electrolyte depth increases. For this reason, data that has been analyzed using SPIDER can be checked for deviations from nominal values of these parameters in order to diagnose any errors that may have occurred in setting up the experiment or taking data. Furthermore, the advection term $c_A$ was only marginally affected as these parameters were varied, so for the purpose of troubleshooting experimental data, this term can largely be ignored.

With each global parameter that is varied, SPIDER reliably returns the same governing equation, the modified Navier-Stokes \refeqs{Q2DNS}, however it is not immediately obvious why this should be the case. The starting candidate model includes terms like $(\nabla\cdot\bfU)\bfU$, and while it is true that this fluid is incompressible and thus $\nabla\cdot\bfU=0$, this is only true of the 3D fluid, whereas the candidate term is only taking a 2D divergence. It is possible that as the fluid depths become large enough, 3D effects could manifest as non-negligible 2D divergence terms in the parsimonious model \cite{akkermans_2008a, akkermans_2008b}, but this is not the case in the range of parameters explored here. In other words, it may not be true in general that the same model will always be returned, but the appearance of different terms could be an indication of new significant physical phenomena arising from changes in system parameters \cite{gurevich_2021}.

We should not be surprised that SPIDER returns \refeqs{Q2DNS} as a parsimonious model, since it is precisely the same as the depth-averaged model identified from a first principles approach which assumes a separable 3D flow ${\bf V}(x,y,z,t) = P(z){\bf U}(x,y,t)$ \cite{suri_2014} (find details in the Supplementary Notes). The SPIDER coefficients are in some cases significantly larger than the first principles predictions, a disparity that holds true over a large range of Re \cite{reinbold_2021}. The discrepancies between the SPIDER and first principles coefficients should also be expected, as the ansatz made in \cite{suri_2014} assumes a stationary flow at very low Re, and thus a velocity profile $P(z)$ that is uniform in $x,\,y,$ and $t$, while the experiments performed are time-dependent by necessity for linear independence in the SPIDER algorithm. We have no reason to assume that the first principles model should even be valid in the Re regime in which these experiments are conducted, but the results of our regression at high Re imply that the form of the first principles model does indeed apply even with significant driving, as no new terms are returned by SPIDER.

While we cannot directly compare the values of the coefficients from SPIDER to those from the first principles model, we can compare the trends in their functional behaviors. As the global parameters change, so too does the profile $P(z)$ predicted in \cite{suri_2014}, so we compute the first principles predictions for the profile and thus the coefficients $c_{\text{FP}}$ in the same range of parameters as the experimental data, as shown by the red dotted curves in each of \reffig{c_hc}-\ref{fig:c_T}. These results are then scaled by a constant factor that minimizes a least-squares cost function with the SPIDER results,
\begin{equation}\label{eq:costfn}
    Z_{ij}(m_{ij}) = \sum_{k=1}^{N_j} (c_{ijk}-m_{ij}c_{ijk,\text{FP}})^2
\end{equation}
where the index $i \in \{A,V,F\}$ labels the coefficients, $j\in\{h_c,h_d,T\}$ labels the global parameter being altered, and $k$ sums from 1 to the number of experiments performed for each parameter, $N_j$. $\tilde m_{ij}$ is the unique multiplicative factor for each coefficient and parameter choice that minimizes this cost function, 
\begin{equation}\label{eq:multiplier}
    c_{ij,\text{FP}} \rightarrow \tilde m_{ij}c_{ij,\text{FP}} \;\; \text{where} \;\; \tilde m_{ij} = \argmin_{m_{ij}} Z_{ij}(m_{ij}).
\end{equation}

This rescaling allows us to directly compare the functional relationships between the global parameters and the SPIDER and first principles coefficients, as shown by the blue curves in \reffig{c_hc}-\ref{fig:c_T}. Remarkably, both the machine learning and first principles method predict not only the same qualitative trends, but quantitative agreement can be found in all three of the coefficients across the gamut of parameters tested in this study. This serves as an exceptional validation of the predictive ability of the first principles model well beyond the driving regime in which it is formulated, despite the breakdown of many of the assumptions upon which it is built.

In this article we have demonstrated that machine learning can be a powerful tool to identify drifting global parameters in a dynamic system. We have provided evidence that changes in particular parameters influence the learned governing model in predictable and distinct ways. By observing the trends in the variations of this model, we have shown that in principle one can determine uniquely which global parameter shift is responsible for these variations. Moreover, the trends observed from data aligned qualitatively with those predicted by a first principles analysis, and after a rescaling displayed identical quantitative trends as well. While this demonstration was performed in the context of a turbulent fluid flow, the method used is entirely general and can be applied to any type of quantitative data, from climate reports to test scores to blood diagnostics. 


\section{Methods}

\subsection{Experimental Methodology}
In order to identify the effects of varying the fluid properties on the model and its coefficients, we run the experiment for 30 minutes, slowly varying each parameter. This length of time is about 60 times longer than the characteristic timescale of the flow at the driving used for these experiments, Re $\approx$ 20 \cite{suri_2017}. A 2D velocity field time series is reconstructed from video of each trial, and this data is used by SPIDER to determine the parsimonious model. 

To characterize the effects of a varying electrolyte depth, the experiment is constructed with an electrolyte layer with depth $h_c = 4.0$ mm. After each data set, the fluid is drained via syringe to lower the depth by 0.2 mm until a final depth of 2.0 mm is reached. This range of fluid depths used were selected based on the stability of the bi-layer. To maintain a constant current density $J$ and thus a constant driving force, the current is increased to compensate for the decrease in the cross-sectional area of the fluid. 

The same syringe draining methodology is used to measure the effects of changing the dielectric depth. However, since siphoning off the dielectric brings the electrolyte layer closer to the magnet array, the current is decreased at each step to maintain a constant Re while the magnetic field strength increases. 

To determine the effects of changing the temperature of the working fluids, we first characterize the density and viscosity response of the electrolyte and dielectric to temperature changes (see the Supplementary Notes). Since both fluids become less viscous as temperature increases, the driving current is decreased as temperature is increased to keep a constant Re across trials. The experiment is first established at 16$^\circ$C. After each data set, the temperature is increased by 1.75$^\circ$C and allowed to come to thermal equilibrium via a PID loop, and the next data set is taken, up to 30$^\circ$C. 

The first principles coefficients are explicitly dependent on the fluids' depths, densities, and viscosities, as well as properties of the forcing like the current density and magnet width and field strength, but not temperature. To determine the effect of $T$ on the first principles coefficients, we fit the fluids' viscosities and densities with second-order polynomials (see the Supplementary Notes) to determine values for the first principles coefficients in \reffig{c_hc}-\ref{fig:c_T}. 

\subsection{The SPIDER algorithm}

As detailed in \cite{reinbold_2021}, the SPIDER algorithm starts with a candidate model of terms ${\bf F}_n$ that are deemed physically suitable \emph{a priori},
\begin{align}
 \partial_t{\bf u} & = \sum_n c_n {\bf F}_n[{\bf u}, p, {\bf f}, \nabla{\bf u}, \nabla p, \nabla {\bf f}, \dots] \\
 &= c_1({\bf u}\cdot\nabla){\bf u} + c_2\nabla^2{\bf u} + c_3{\bf u} + c_4u^2{\bf u} + c_5\omega^2{\bf u} \nonumber \\
 & + c_6 (\nabla\cdot{\bf u}){\bf u} + c_7 (\nabla\cdot{\bf u})^2{\bf u} - \rho^{-1}\nabla p + \rho^{-1}{\bf f}  \nonumber.
 \label{eq:u}
\end{align}
Each of these terms is then evaluated on a domain $\Omega_i$ and integrated,
\begin{equation}\label{eq:sp}
 \langle {\bf w}_j, {\bf F}_n \rangle_i = \int_{\Omega_i} {\bf w}_j\cdot{\bf F}_n d\Omega,
\end{equation}
where $\td \Omega = \td x \td y \td t$. The analytic weight functions ${\bf w}_j$ are specially chosen to allow derivatives in the terms ${\bf F}_n$ to be moved off of noisy experimental data through integration by parts. We evaluate \refeqs{sp} at various values of $i$ and $j$ and stack the results to form vectors ${\bf q}_n$, where $n=0$ corresponds to $\partial_t {\bf u}$. We then form the linear system 
\begin{equation}
    Q{\bf c} = {\bf q}_0,
\label{eq:c}
\end{equation}
where ${\bf c}=[c_1,\cdots,c_N]^T$ and $Q=[{\bf q}_1\ \cdots\ {\bf q}_N]$. This over-determined system is solved by iteratively finding the coefficients ${\bf c}$, removing those terms whose magnitude lies below a threshold, and solving again until a parsimonious model is returned. The model is describes the data well when the residual $\eta$ is low, 
\begin{equation}\label{eq:eta}
 \eta = \frac{\| Q{\bf c} - {\bf q}_0 \|}{\max_n\| c_n{\bf q}_n \|},
\end{equation}
In all data sets, the same parsimonious model was recovered with low residual. As described in \cite{reinbold_2021}, the pressure field and forcing can be reconstructed from this data, and thus values for $c_f$ and $c_p$ can be derived, but this is beyond the scope of this study. 

Although the identified terms and their numerical coefficients are robust to changes in Re in the range 20-40 for which the flow is turbulent \cite{tithof_2017,reinbold_2021}, we choose to fix Re to isolate the effects of only one global parameter. Since changing the geometry and characteristics of the fluids inherently changes the forcing experienced, measures are taken to ensure that Re is held constant throughout each trial.

\section{Supplementary Material}

\subsection{Kolmogorov-Like Experimental Setup}

\begin{figure}[htb]
    \centering
    \includegraphics[width=0.95\columnwidth, valign=c]{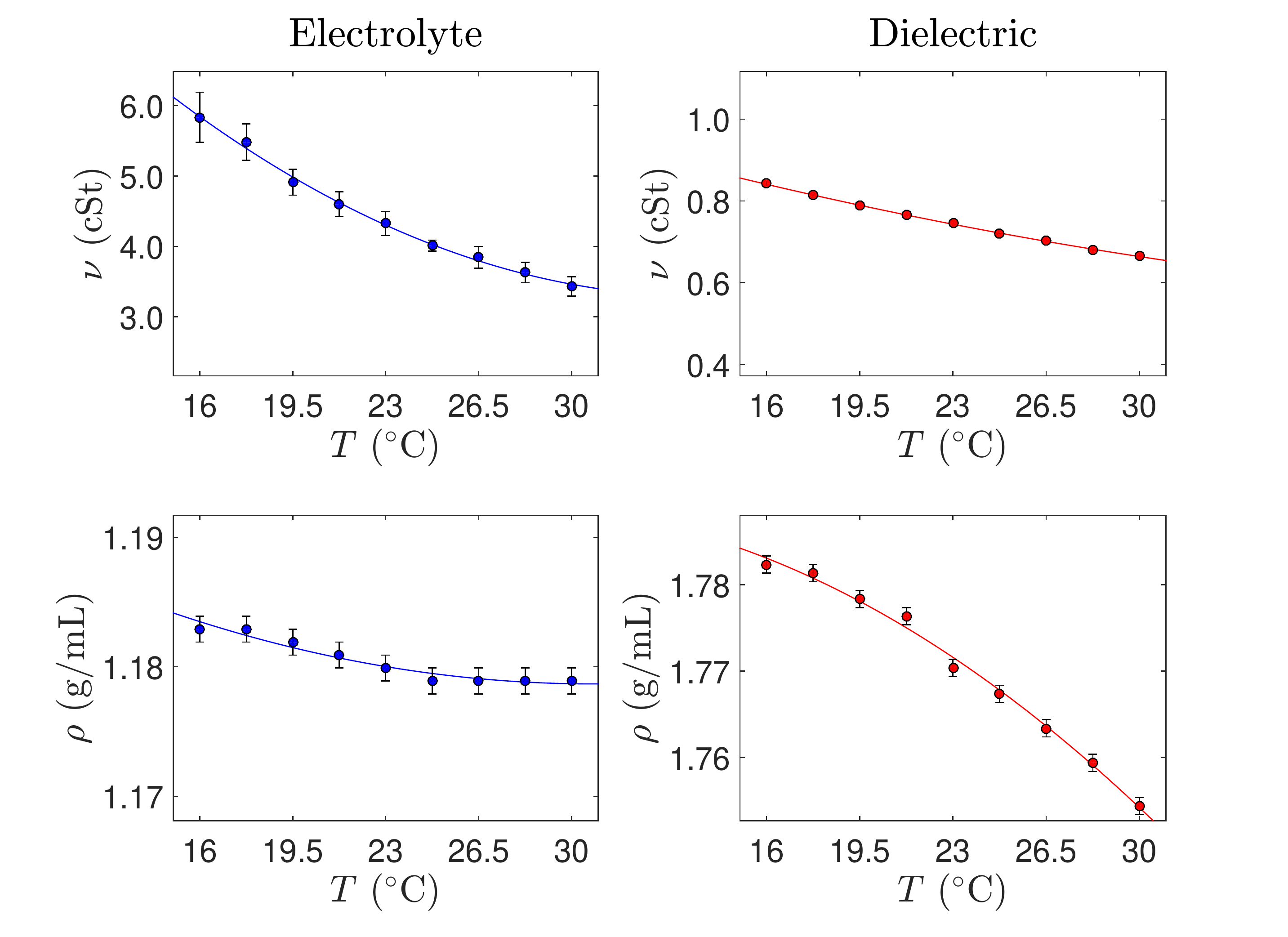}
    \caption{{\bf Fluid properties.} Both the dielectric and electrolyte have temperature-dependent properties. Densities $\rho$ were measured with hydrometers, while kinematic viscosities $\nu$ were measured with viscometers. All fit curves are second order polynomials, and error bars represent standard deviations.}
    \label{fig:nu_rho}
\end{figure}

The experimental apparatus is similar to that used in Ref. \cite{tithof_2017}.
In a rectangular container, a flow is produced in a shallow, immiscible bi-layer of conducting electrolyte (copper sulfate and glycerine solution) and dielectric (perfluorooctane). 
The nominal kinematic viscosities of these fluids are $\nu_c = 4.302$ cSt and $\nu_d = 0.743$ cSt, and their densities are $\rho_c = 1180$ kg/m$^3$ and $\rho_d = 1772$ kg/m$^3$.
As properties are functions of temperature, a Cannon-Fenske viscometer was used to measure the viscosity and a hydrometer was used to determine the density of each fluid at the temperatures used in these trials. 
The results of these measurements are shown in \reffig{nu_rho} along with second order polynomial fits that were used to determine the temperature dependence of the coefficients in the depth averaged model.
Both layers have a nominal thickness of $h_c = h_d = 3$ mm, and the dimensions of the container are $L_x=17.8$ cm $\times$ $L_y=22.9$ cm. 
The container sits on a thermal reservoir, which limits temperature fluctuations to $0.1 ^\circ$C about the set temperature, with nominal working temperature 23 $^\circ$C. 
The dielectric layer acts as a lubricant, preserving the desired quasi-two-dimensionality within the electrolyte layer.
The bottom of the container does still create a no-slip condition, which creates a varying flow velocity profile in the vertical direction which depends on the fluid properties, as detailed in the next section.

\begin{figure}[htb]
    \centering
    \includegraphics[width=0.9\columnwidth, valign=c]{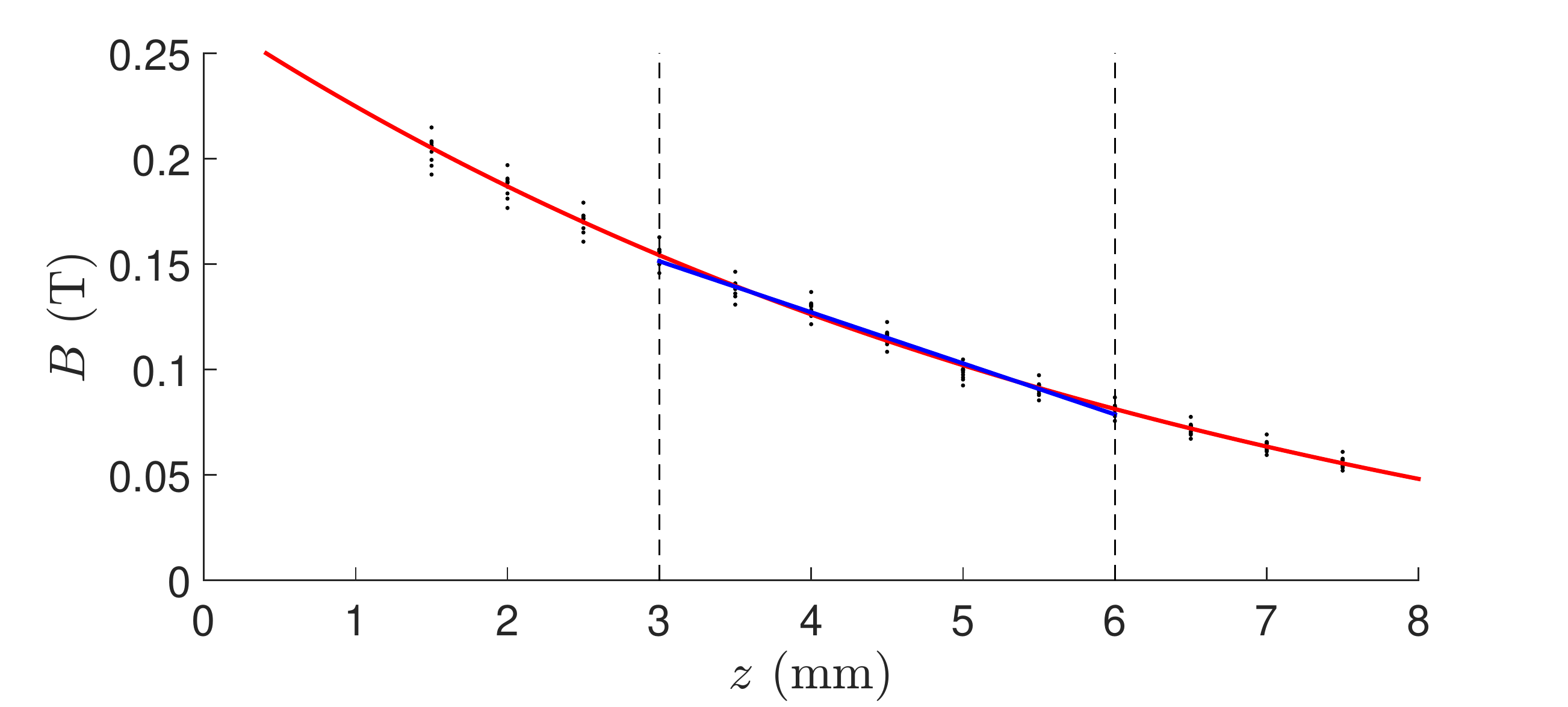}
    \caption{{\bf Magnetic field measurements.} The measured magnetic field $z$ component at varying heights $z$ above the magnets (black points, standard deviation error bars are smaller than the symbols). The red curve is an exponential nonlinear least-squares fit, and the blue line is a linear fit within a typical electrolyte layer, delineated by vertical dotted lines. Within any electrolyte layer thinner than 4 mm, the linear fit has $r^2 > 0.98$, demonstrating that the exponential decay model is not necessary in the first-principles derivation.}
    \label{fig:Bfield}
\end{figure}

A pair of electrodes generates a constant direct current with density ${\bf J}=J\hat{y}$ through the electrolyte layer. 
An array of 14 permanent magnets of width $w=1.27$ cm placed beneath the container maintains a constant magnetic field $\mathbf{B}$ that is near-sinusoidal in the center of the domain. 
The interaction of the current density and magnetic field produces a Lorentz force ${\bf J}\times {\bf B}$ that drives the flow. 
The $z$-component of the magnetic field has been measured at the center of each of the middle 10 magnets at 13 equally-spaced heights in order to determine the decay rate of the field, as shown \reffig{Bfield}. While the field decays exponentially, within the small range of distances spanned by the electrolyte layer, the decay is well-approximated by a linear function $B_z = B_1z + B_0$.
These measurements were used to determine the velocity profile within the fluid bi-layer.

Fluorescent polyester particles with radius $\approx$ 25 $\upmu$m seed the electrolyte-dielectric interface and are illuminated by LED strips placed alongside the apparatus. A camera above takes video data sets, and particle image velocimetry (PIV) is then used to measure 2D velocity fields that quantify the horizontal flow \cite{prana}. 

The strength of the flow is characterized by the Reynolds number Re$=\bar{u}w/c_2$, where $\bar{u}$ is the RMS velocity within the central $8w \times 8w$ region of the domain, and $c_2$ is the characteristic depth-averaged viscosity as predicted by the first-principles model, which allows direct comparison with the results of previous studies of this experimental system \cite{suri_2014, tithof_2017, suri_2017a, suri_2018, suri_2019, reinbold_2021}. 
The vertical ($z$) component of the flow is negligibly small for Re $\lesssim 50$, so the horizontal flow can be considered divergenceless \cite{suri_2014}.

Each data set consists of a 30-minute time series of the PIV-reconstructed $x$ and $y$ components of the velocity field on a uniform grid within the flow domain with temporal resolution $\Delta t=1$ s. 
The characteristic time scale $\tau$ of the flow varies with Re.
At the Re $\approx$ 20 tested in this study, the flow is aperiodic, with $\tau \approx 30$s \cite{suri_2017a}. 
The spatial resolution of the data is about 8.4 grid points per magnet width $w$, which is the characteristic length scale of the flow. 

\subsection{Q2D Depth-Averaged Model}
The velocity field $\bfV$ generated by the experimental setup obeys the incompressible three-dimensional (3D) Navier-Stokes equations, however confinement in the vertical direction and the limitations of measurement techniques inspire the creation of a Q2D model for our shallow fluid flow.
To derive the model from a first-principles approach, we follow the methodology of \cite{suri_2014} and assume that the full 3D flow can be decomposed as the product of a vertical velocity profile and a 2D velocity field, $\bfV(x,y,z,t) = P(z)\bfU(x,y,t)$. The validity of this assumption is discussed in \cite{satijn_2001}.

Plugging this decomposition into the full 3D Navier-Stokes equation, we retrieve an equation in the horizontal direction, 
\begin{equation}\label{eq:3DNS_hor}
     \rho P \partial_t \bfU + \rho P^2\bfU\cdot \nabla \bfU = - \nabla p + \mu P \nabla^2 \bfU + \mu \bfU \partial_z^2 P + \bff.
\end{equation}
where $\rho,\, \mu,$ and $\bff$ are the fluid's density, dynamic viscosity, and driving force respectively, and $\nabla$ acts only in the horizontal direction.

Integrating \refeqs{3DNS_hor} through the $z$ direction, dividing by the prefactor on the time derivative term, and rearranging, we obtain the Q2D governing equation for the horizontal velocity,
\begin{equation}\label{eq:Q2DNS}
     \partial_t \bfU = c_1 \bfU\cdot \nabla \bfU + c_2 \nabla^2 \bfU + c_3 \bfU + c_f \bar{\bff} - c_p\nabla p
\end{equation}
where
\begin{equation}\label{eq:coeff_def}
\begin{gathered}
    c_1 = -\frac{\int_0^h \rho P^2 \td z}{\int_0^h \rho P \td z}, \qquad
    c_2 = -\frac{\int_0^h \mu P \td z}{\int_0^h \rho P \td z}, \\[0.8em]
    c_3 = -\frac{[\mu P']_{z=0}}{\int_0^h \rho P \td z}, \qquad
    c_p = c_f = \frac{h}{\int_0^h \rho P \td z}
\end{gathered}
\end{equation}
and $\bar{\bff}$ is the depth-averaged forcing.

To solve for the velocity profile $P$ we again follow the methodology in \cite{suri_2014}, which derives a piecewise profile in the conducting (c) layer where $h_d < z < h_d+h_c$
\begin{equation}\label{eq:profile_c}
    P = Ce^{\kappa z} + De^{-\kappa z} + \frac{JB_1}{u_0\mu_c\kappa^2}z + \frac{JB_0}{u_0\mu_c\kappa^2} 
\end{equation}
and in the dielectric (d) layer where $0 < z < h_d$:
\begin{equation}\label{eq:profile_d}
    P = Ee^{\kappa z} + Fe^{-\kappa z} 
\end{equation}
where $\kappa = \pi/w$ is the spatial frequency of the magnet array and approximates the frequency of the near-sinusoidal Lorentz force, $J$ is the current density, and the magnetic field $z$ component is modelled as linearly decaying, $B_z = (B_1z + B_0)\sin(\kappa y)$, as in \reffig{Bfield}.

With boundary conditions
\begin{equation}\label{eq:prof_bc}
    \begin{gathered}
    P(0) = 0,\qquad P(h_d^-) = P(h_d^+), \\
    \mu_d P'(h_d^-) = \mu_c P'(h_d^+), \qquad 
    P'(h_d+h_c) = 0,\\ 
    P(h_d+h_c) = 1
    \end{gathered}
\end{equation}
one can solve for the five unknowns $C,\, D,\, E,\, F,$ and $u_0$ and thus find analytical expressions for the Q2D coefficients, which are too long to include in this manuscript. These coefficients are functions of the depths, densities, and viscosities of both the electrolyte and the dielectric. At nominal values of all of these parameters as described in the previous section, the values of the numbered coefficients are $c_1 = 0.819,\,c_2 = 2.88$ mm$^2$/s, and $c_3 = 0.0630$ s$^{-1}$. The value of $c_f$ and $c_p$ is simply the inverted depth-averaged density, but since the pressure and forcing cannot be directly measured, we will not be comparing the numerical value of this coefficient.

\subsection*{Data availability}
Data sets containing velocity fields are available from the Open Science Framework at \href{https://doi.org/10.17605/OSF.IO/BHU6M}{https://doi.org/10.17605/OSF.IO/BHU6M}. 

\subsection*{Code availability}
MATLAB codes used to identify the governing equations can be found on GitHub at \href{https://doi.org/10.5281/zenodo.4653308}{https://doi.org/10.5281/zenodo.4653308}. 
Any other requests should be made to the corresponding author.

\subsection*{Acknowledgements} This material is based upon work supported by NSF under Grants No. CMMI-1725587 and CMMI-2028454.

\subsection*{Author contributions}

L.M.K. was responsible for performing fluid flow experiments, data acquisition, PIV analysis, and SPIDER.
R.O.G. was responsible for concept design. 
M.F.S. was responsible for experimental and research design.
All authors were involved in the preparation of the manuscript, read and approved the final version.

\subsection*{Competing Interests}
The authors declare no competing interests.

\bibliographystyle{unsrt}
\typeout{}
\bibliography{ml}

\end{document}